\documentclass[11pt, twoside]{article}
\usepackage{amsmath}
\usepackage{float}
\usepackage{amsthm}
\usepackage{amssymb}
\usepackage{graphicx}
\usepackage{amsmath}
\usepackage{mathtools}
\usepackage{amsfonts}
\usepackage{amssymb}
\usepackage{graphicx}
\usepackage{longtable}
\usepackage{adjustbox}
\usepackage{lipsum}
\usepackage{abstract}
\usepackage{hyperref}
\usepackage[ruled,vlined]{algorithm2e}
\usepackage[utf8]{inputenc}
\usepackage[english]{babel}

\usepackage[square,numbers]{natbib}
\usepackage{adjustbox}
\usepackage{lipsum}
\usepackage[font=scriptsize]{caption}
\usepackage{graphicx,type1cm,eso-pic,color}

\makeatletter
          \AddToShipoutPicture{
            \setlength{\@tempdimb}{.5\paperwidth}
            \setlength{\@tempdimc}{.5\paperheight}
            \setlength{\unitlength}{1pt}
            \put(\strip@pt\@tempdimb,\strip@pt\@tempdimc){
        \makebox(0,0){\rotatebox{55}{\textcolor[gray]{0.85}
        {\fontsize{3cm}{3cm}\selectfont{JISA-DRAFT}}}}
            }
        }
\makeatother

\numberwithin{equation}{section}

\begin{document}
\setcounter{page}{1}
\thispagestyle{empty}
\markboth{}{}

\pagestyle{myheadings}
\markboth{Journal, Indian Statistical Association}{Real-Time Monitoring and Forecasting of COVID–19 Cases}

\date{\today}


\noindent  Journal of the Indian Statistical Association\\
Vol.53 No. 1 \& 2, 2015,

\vspace{.1in}

\baselineskip 20truept

\begin{center}
{\Large {\bf Real-Time Monitoring and Forecasting of COVID–19 Cases using an Adjusted Holt–based Hybrid Model embedded with Wavelet–based ANN}} \footnote{Received : January 2021}
\end{center}

\vspace{.1in}

\begin{center}
{\large {\bf Agniva Das}} \\
{\large {\it Department of Statistics, Faculty of Science, The Maharaja Sayajirao University of Baroda, Vadodara - 390002}} \\
{\large {\bf Kunnummal Muralidharan}}\\
{\large {\it Department of Statistics, Faculty of Science, The Maharaja Sayajirao University of Baroda, Vadodara - 390002}}\\
\end{center}

\vspace{.1in}
\baselineskip 18truept

\begin{abstract}

Since the inception of the SARS - CoV - 2 (COVID - 19) novel coronavirus, a lot of time and effort is being allocated to estimate the trajectory and possibly, forecast with a reasonable degree of accuracy, the number of cases, recoveries, and deaths due to the same. The model proposed in this paper is a mindful step in the same direction. The primary model in question is a Hybrid Holt's Model embedded with a Wavelet-based ANN. To test its forecasting ability, we have compared three separate models, the first, being a simple ARIMA model, the second, also an ARIMA model with a wavelet-based function, and the third, being the proposed model. We have also compared the forecast accuracy of this model with that of a modern day Vanilla LSTM recurrent neural network model. We have tested the proposed model on the number of confirmed cases (daily) for the entire country as well as 6 hotspot states. We have also proposed a simple adjustment algorithm in addition to the hybrid model so that daily and/or weekly forecasts can be meted out, with respect to the entirety of the country, as well as a moving window performance metric based on out-of-sample forecasts. In order to have a more rounded approach to the analysis of COVID-19 dynamics, focus has also been given to the estimation of the Basic Reproduction Number, $R_0$ using a compartmental epidemiological model (SIR). Lastly, we have also given substantial attention to estimating the shelf-life of the proposed model.
It is obvious yet noteworthy how an accurate model, in this regard, can ensure better allocation of healthcare resources, as well as, enable the government to take necessary measures ahead of time.
\\ Entire code available at  \href{https://github.com/dasagniva/wbann}{the GitHub repository (click on text)}.

\end{abstract}

\vspace{.1in}

\noindent  {\bf Key Words} : {\it ARIMA, artificial neural networks, COVID-19, exponential smoothing, forecasting, Holt's model, hybrid model, time series, wavelet transform, moving window}

\section{Introduction}

As per the report by World Health Organization as on January $10^{th}$, 2021, there have been  confirmed 88,828,387 cases of COVID-19, including 1,926,625 deaths, globally. India, alone, has contributed to roughly 11.78\% of confirmed cases and 7.85\% of reported deaths, globally. India's experience with the pandemic is critical due to India's high population count as well as density. Nearly 50\% of the total number of confirmed cases of COVID-19 in India is distributed among 5 individual states, which have been termed "hotspots" due to the high counts of daily confirmed cases. Hence, it becomes imperative in a sense, to not only identify the hotspots, but to determine any sharp increase in the counts in the near future. As of January $10^{th}$, 2021, Maharashtra has the highest number of confirmed cases contributing to roughly 18.81\% of the nationwide count of confirmed cases, followed by Andhra Pradesh, with 8.44\% , Tamil Nadu with 7.89\%, Karnataka with 8.85\%, Chhattisgarh with 2.76\%, Kerala with 7.77\%. \\ \\
Forecast models for epidemiological dynamics of the novel coronavirus pandemic has been common recently. Most models, however, have been developed with scarce data. This entails a severe problem, with respect to the compromise(s) the researcher/modeller must make to maintain model sparsity, as well as to avoid the risk of overfitting at the same time. With more data being available as time passes, more and more complex models are being explored. Here, we have explored some of the models that have shown promise in terms of out of sample predictions. We have also employed a simple yet effective technique to select the best possible model amongst those discussed in this paper, using out of sample forecasts as the point of focus for model selection.\\ \\
\cite{Aladag2009} proposed a hybrid model which was applied to Canadian Lynx data consisting of the set of annual numbers of lynx trappings
in the Mackenzie River District of North–West Canada for the period from 1821 to 1934, and we see that their given model gives the best forecasting accuracy.
\cite{Aladag2009} states in recent years, artificial neural networks (ANNs) are being used extensively for forecasting time series in the literature. Although we can to model both linear and nonlinear structures in time series by using neural networks, ANNs are unable to handle both structures .
In \cite{Aminghafari2007} paper deals with wavelets in time series, with a focus on statistical forecasting purposes. In the more recent approaches wavelet decompositions are being used in order to deal with non-stationary time series data.
\cite{Benaouda2006} proposes a wavelet multi-scale decomposition-based autoregressive approach to predict 1-h ahead load based on historical electricity load data. The course of action taken here is based on a multiple resolution decomposition of the signal which makes use of the non-decimated Haar à trous wavelet transform. The benefit is that, it takes into account the asymmetric nature of the time-varying data.
\cite{Boldog2020} developed a computational tool to assess the risks of novel coronavirus outbreaks outside of China. In the paper they estimate that the risk of a major outbreak in a country depends on parameters such as, the hike in the cumulative number of cases in mainland China just outside the closed areas; the travel connection of the destination country with China, which also covers baseline travel frequencies, the consequences of travel restrictions, and the usefulness of entry screening at destination; and the efficacy of control measures in the destination country.
\cite{Chakraborty2019} have proposed a hybrid model which combines ARIMA and NNAR model to combine both the linearity and non-linearity in the datasets. The ARIMA model used excludes out linear tendencies in the data and passes on the residual values to the NNAR model.
\cite{Fanelli2020} proposes an analysis of day-lag maps points to some uniformity in the spreading of the epidemic. A simple mean-field model is used by them to provide a quantitative picture of the pandemic spreading, and especially the height and time of the peak of confirmed infected individuals.
The concern of \cite{Fay2007} is with the use of an exogenous system wherein a model is required to forecast time series using a causal input. A new approach has been used in which the wavelet transform is given to both the dependent time series and causal input.
\cite{Jung2020} suggests the exported cases of 2019 novel coronavirus (COVID-19) infection which has been confirmed outside China provides an excellent opportunity to provide an estimate of the cumulative confirmed cases, fatality risk (cCFR) in mainland China.
In \cite{Khashei2019}, two types of hybrid models have been used for predicting the stock prices which have been used to introduce a more reliable series hybrid model. They have used ARIMA and MLPs for creating series hybrid models.
\cite{Kucharski2020} proposes merging a mathematical model of the SARS-CoV-2 transmission by using four sets of data which have been collected from Wuhan. They also estimated how transmission in Wuhan varied between December, 2019, and February, 2020. Also using these estimates to assess the potential for sustained human-to-human transmission that occur in locations outside Wuhan if cases were introduced.
\cite{Chakraborty2020} have presented a hybrid approach which is based on ARIMA model and an impactful Wavelet-based forecasting model which have been used to generate short-term forecasts for the number of daily confirmed cases.
\cite{Petropoulos2020} have described a timeline of forecasts which have been used to propose planning and decision making as well as providing objective forecasts for the confirmed cases of COVID-19.
\cite{Roosa2020} uses phenomenological models that have already been developed and used for previous outbreaks in order to generate short-term forecasts of the cumulative number of confirmed reported cases in China's Hubei province.
\cite{Wu2020} provides an estimate of the extent of the Corona epidemic in Wuhan based on the basis of the number of cases from Wuhan to cities outside of China and have forecasted the extent and impact of the domestic and global health risks, accounting for social and non-pharmaceutical prevention interventions.
\cite{Zhang2003} proposes a hybrid method that makes use of both ARIMA and ANN models and proposes the use of ARIMA and ANN models in linear and nonlinear modeling.
In this study \cite{Zhuang2020} have used the data on air travel and the number of cases from Iran imported into various Middle Eastern regions and then provide an estimate of the number of COVID-19 cases in Iran.
\cite{Liu2020a} gives a retrospective study of patients with new coronavirus pneumonia (COVID-19) who were previouslyhospitalized in Hainan Provincial People's Hospital from January 15, 2020 to February 18, 2020.
In this paper \cite{Khrapov2020} proposes a mathematical model for the coronavirus COVID-19 epidemic development and is performed for China using the epidemiological SIR model.
In this study, \cite{Zhong2020} have presented an early prediction of the epidemic of the 2019-nCoV based a simplified SIR model. The rationality of the available epidemiological data was analyzed so as to obtain an estimation of the key parameter, i.e., infection rate.
In this paper, \cite{Ribeiro2020} suggests an autoregressive integrated moving average (ARIMA), cu- bist regression (CUBIST), random forest (RF) and other methods which have been evaluated for time series forecasting. \\ \\
One daily forecast has been demonstrated for the number of confirmed cases of contraction of the novel coronavirus in India, including 6 hotspot States.

\section{Real-Time Monitoring \& Forecasting}
The advent of several new open-source data banks have provided significant easement in terms of sourcing real-time data. This, generally paves the way for forecasting models to render themselves to a state of continuous improvement in terms of model accuracy.\\ \\
Here, we have sourced the data from \href{https://www.covid19india.org/}{covid19india.org}$^{1}$, a crowdfunded operation that has scraped data from various official sources mentioned \href{https://telegra.ph/Covid-19-Sources-03-19}{here}$^{2}$.

\subsection{Dataset}
Although individual univariate time series datasets are readily available for India and each of its states, we have only considered including the data streams for the total cases (confirmed) in India, and six of its hotspot states, namely, Maharashtra, Tamil Nadu,  , Karnataka, Chhattisgarh, Kerala  and Andhra Pradesh. Data for all the other states in the country can be viewed \href{https://www.covid19india.org/}{here}$^{3}$. Only the cases, be it confirmed, or recovered or deceased, from March 14, 2020 to January 10, 2021 have been considered for all states as well as the entire country.\\ \\
A total of 272 observations have been considered for each univariate time series dataset. As of now, diminishing nature has been (very recently) observed in the epidemic curve, hence, we have focused largely on trended and non-seasonal models for the forecasting purpose. \\ \\
\textbf{Note:} All the models discussed in this paper are of the non-intervention type and assume that the trend continues indefinitely in the future.

\subsection{Proposed Model}
We have adopted a hybrid forecasting approach in the form of an amalgamation of the basic Holt's model and a wavelet based artificial neural network. As there is no seasonality observed in any of the datasets, the Holt's model has been selected for the purpose of forecasting. Additionally, the hybridization of the model enables us to minimize to great extent, the deficiencies of the single time series models. We have adopted a methodology similar to that of \cite{Chakraborty2020}, which we shall discuss in the forthcoming subsections.

\subsubsection{Holt's Model}
Owing to the lack of seasonality observed in the number of confirmed, recovered or deceased cases in India and all of the states taken into consideration, and the obvious notion of recent data points dominating the future estimates, the choice of Holt's Trend Corrected Exponential Smoothing Model, more popularly known as the Double Exponential Smoothing Model seems plausible. The two parameter exponential smoothing model with additive trend, proposed by Holt (1957), is comprised of two components, namely $\alpha$, the smoothing constant used to control speed of adaptation to local level and $\beta$, another smoothing constant, introduced to control degree of local trend carried through to multi-step ahead forecast periods. The recursive form of the equation can be expressed as
\begin{center}
\textbf{Level Equation:} $L_t = \alpha Y_t + (1 - \alpha)(L_{t-1} - T_{t-1})$ \\
\textbf{Trend Equation:} $T_t = \beta (L_t - L_{t-1})+(1-\beta)T_{t-1}$\\
\textbf{Forecast Equation:} $\hat{Y}_{t+h|t} = L_t+hT_t$  \\
\end{center}
where, $L_t$ denotes an estimate of the level of the time series at time $t$, $T_t$ denotes the estimate of the slope of the trend at time $t$, $\hat{Y}_{t+h|t}$ denotes the forecasted value of the time series at time $t+h$ based on the obeservations obtained till time $t$, $\alpha$ denotes the data smoothing factor and $\beta$ denotes the trend smoothing factor.
\\ The choice of Holt's Double Exponential Smoothing Method of forecasting over ARIMA can be attributed to two major arguments:
\begin{itemize}
\item ARIMA carries with it, the assumptions of stationarity, i. e. errors having zero mean and constant variance independently and identically.
\item There was no seasonality observed in any of the time series considered.
\end{itemize}
One may point out that ARIMA may provide a better fit, but when coupled with the wavelet-based ANN, the Holt hybrid fares substantially better, as compared to its ARIMA counterpart discussed in \cite{Chakraborty2020}.

\subsubsection{Wavelet-based ANN}
ANN as a statistical method is generally non-parametric, self-adaptive, nonlinear and data-driven. It is a powerful modelling tool, especially when the underlying data relationship is unknown. Additionally, the freedom from restrictive assumptions like linearity which lends to mathematical models rendering themselves useful, is one of its main attractions, along with its high learning capacity. Providing an implicit functional representation of time often renders the ANN useful for time series data, bestowing static neural networks like the multi-layer perceptron with dynamic properties.To cite one example, \cite{Anjoy2017}, have successfully used the Time-delay neural network (TDNN) to build short-term memory into the structure of the neural network, through time delay, implemented at the input layer of the neural network, with a logistic function in its activation layer.
\\
Wavelet transformations are signal processing tools considered to be able to produce both time and frequency information with higher resolution, effectively. For time series data, however, there are two approaches to the creation of wavelet based neural networks.
\begin{itemize}
\item \textbf{Wavelet-Based ANN (WBANN) approach:} Transforming (decomposing) the original signal and representation in different domains which are flexible in terms of analysis and processing.
\item \textbf{Wavelet Neural Network (WNM) approach:} Choosing a wavelet-based activation function.
\end{itemize}
The approach referred to in this paper is the first, wherein the TDNN has been employed to fit individual detail and approximate a part of the decomposed series. In case of non-linearity, the maximal overlap discrete wavelet transform (MODWT), which is well defined for all sample sizes and shift invariant as well, decomposes the original series into sub-frequencies, thereby allowing the ANN to separately model the details and approximate the part more effectively, hence increasing forecast accuracy. Moreover, wavelet decomposition enables transformation of the original nonlinear pattern into several sub-frequencies thereby, capturing the overall as well as partial features of the series which would be impossible in simple ANN methodology. The Haar filter has been employed because:
\begin{itemize}
\item It is symmetric, so that linear phase can be achieved, unlike other wavelets such as Daubechies, Coiflet, etc.
\item It has one vanishing movement; other wavelets have more vanishing movements.
\end{itemize}
Additionally, the Haar wavelet is memory efficient, exactly reversible without the edge effects characteristic of other wavelets, computationally cheap and does not have overlapping windows. It uses just two scaling and wavelet function coefficients, thus calculates pair wise averages and differences.

\subsubsection{Hybrid Holt-WBANN Model}
In pursuit of individual bias reduction of component models in the case of the COVID-19 dataset, the Holt's model, we propose a hybridization of the same using a wavelet-based ANN. The residuals from the Holt's Model, being oscillatory and periodic in nature, lead to the choice of the wavelet-based ANN to model the remaining series. A plethora of similar hybridizations can be seen in the field of time series forecasting.\\ \\

\cite{Khashei2019} hybridized a combination of ARIMA and MLP's (Multi-Layer Perceptrons) in order to predict stock prices. \cite{Zhang2003} introduced the ARIMA-ANN model so as to fit both linear as well as nonlinear components. \cite{Behnamian2010} used a hybrid metaheuristic consisting of a combination of a particle swarm optimizer and a simulated annealer to obtain optimum model parameter estimates.

In view of the above discussion, and along the lines of the methodology mentioned in \cite{Chakraborty2020}, we propose a novel hybrid Holt-WBANN model which is a two-phase approach:
\begin{itemize}
\item \textbf{Phase-I:} A Holt's model is built to model epidemiological dynamics in the form of a time series of the number confirmed/deceased COVID-19 crisis. Out-of-sample forecasts are also generated.
\item \textbf{Phase-II:} The residuals of the model from Phase-I are remodeled using a wavelet-based ANN.
\end{itemize}

Here, the WBANN models the remaining autocorrelations which the Holt's model could not. This renders the proposed approach be looked upon as an error remodelling approach where Holt's Model plays the role of a base model and the WBANN remodels the error series, thereby generating more accurate forecasts.

\begin{algorithm}[H]
\SetAlgoLined
\begin{enumerate}
\item \textbf{Given:} Time series of length n\\
 \textbf{Input:} Training Dataset (Daily COVID-19 confirmed/deceased cases)
\item Fit the Holt's Model iteratively, obtaining the best values for $\alpha$ and $\beta$ as mentioned in section 2.2.1.
\subitem Save the residuals $e_t$ from the Holt's Model
\item Train the residual series $e_t$ generated by the Holt's Model by the WBANN Model as described in section 2.2.2
\subitem Select the decomposition level $W_L=[log(n)]$, and choose the boundary to be periodic.
\subitem Generate fitted values $\hat{e}_t$ and forecasted values $\hat{e}_{t+i|t}, i=1,2,3,...,h$ using WBANN.
\item Combine fitted and forecasted values from both models to form fitted and final forecasts: \\
\textbf{Final Fitted:} $\hat{y}_{t_{hybrid}}=\hat{y}_{t_{Holt}}+\hat{e}_{t_{WBANN}}$ \\
\textbf{Final Forecasted:}$\hat{y}_{t+h|t_{hybrid}}=\hat{y}_{t+h|t_{Holt}}+\hat{e}_{t+h|t_{WBANN}}$
\end{enumerate}
 \caption{\textbf{Proposed Hybrid Holt-WBANN Model}}
\label{algo:model}
\end{algorithm}

\subsubsection{Adjustment factor for Real-Time Monitoring}
Judging from the previous discussions, we can safely imply that as the proposed model is of the non intervention type, it should only be considered for surveillance / monitoring purpose. It does not take into consideration factors like vaccine availability in the market, government interventions, etc. although, in case of most models, like ARIMA, any intervention occurring in the past may well influence the future predictions depending upon the choice of model parameters, whereas the Holt's model simply assigns most weightage to the most recent values, thereby rendering past interventions less influential with time.\\
As our focus is on monitoring, it is imperative that bulk forecasting is a common problem. There also remains the classic problem of the constant sum of multiple time series. We have developed a simple approach for dealing with this problem. \\
Consider a multiple time series ${y_t}^{(i)}, i=1,2,...,n$ denoting the number of confirmed/deceased cases in the $i^{th}$ state at time $t$ and let $Y_t$ denote the number of confirmed/deceased cases in the entire country at time $t$. Then, we simply have $Y_t=\sum \limits_{i=1}^n {{y_t}^{(i)}}$. Now consider the forecasted value ${\hat{y}_{t+1}}^{(i)}, i=1,2,...,n$ of the $i^{th}$ state at time $t+1$. As there is no relation defined within the univariate models between the number cases within states and the number of cases within the entire country, a correction factor becomes a necessity. The adjusted forecasted value, thus becomes
$${\hat{y}^{(i)}}_{{t+1}_{corrected}}= [{\hat{y}_{t+1}}^{(i)}+w_id_{t+1}] I [|Y_t -\hat{Y}_t | \le |\sum \limits_i (y_t^{(i)} - \hat{y}_{t+1}^{(i)})|]+ [\hat{y}_{t+1}^{(i)}] I[|Y_t -\hat{Y}_t | > |\sum \limits_i (y_t^{(i)} - \hat{y}_t+^{(i)})|]$$
 where, $$d_{t+1}=\hat{Y}_{t+1}-\sum \limits_{i=1}^n {\hat{y}_{t+1}}^{(i)}$$ $\hat{Y}_{t+1}$ being the modeled forecast of total number of confirmed cases for the entire country, $I[.] $ being the indicator function, and $w_i$ denotes the proportion of correction allocated to the $i^{th}$ state,
$$w_i =\frac{({{y_t}^{(i)}-{\hat{y}_t}^{(i)})}^2}{{\sum \limits_{i=1}^n {({{y_t}^{(i)}-{\hat{y}_t}^{(i)})}^2}}}, i = 1,2,3,...,n$$
The algorithm provided on \autoref{algo:adj} may introduce some clarity to the protocol being discussed.\\
\\
\textbf{${\textbf{Remark}}^1$:} It may be worthwhile to note that the choice of measure for $w_i$ is critical here. The choice of metric for adjustment here is the sum squared residuals of the last time point for all states. One may choose to take the mean squared residuals for each state over a fixed  time window here, or take exponentially weighted means of sum squared residuals for each state over the entire timeline for this purpose.
\\ \\
\textbf{${\textbf{Remark}}^2$:} The results discussed in this paper are for the non-adjusted case. For results pertaining to the adjusted-models, refer to \autoref{tab:adj}.\\

\subsection{Comparison with similar models}
\begin{figure}[H]
\includegraphics[width=1\linewidth]{"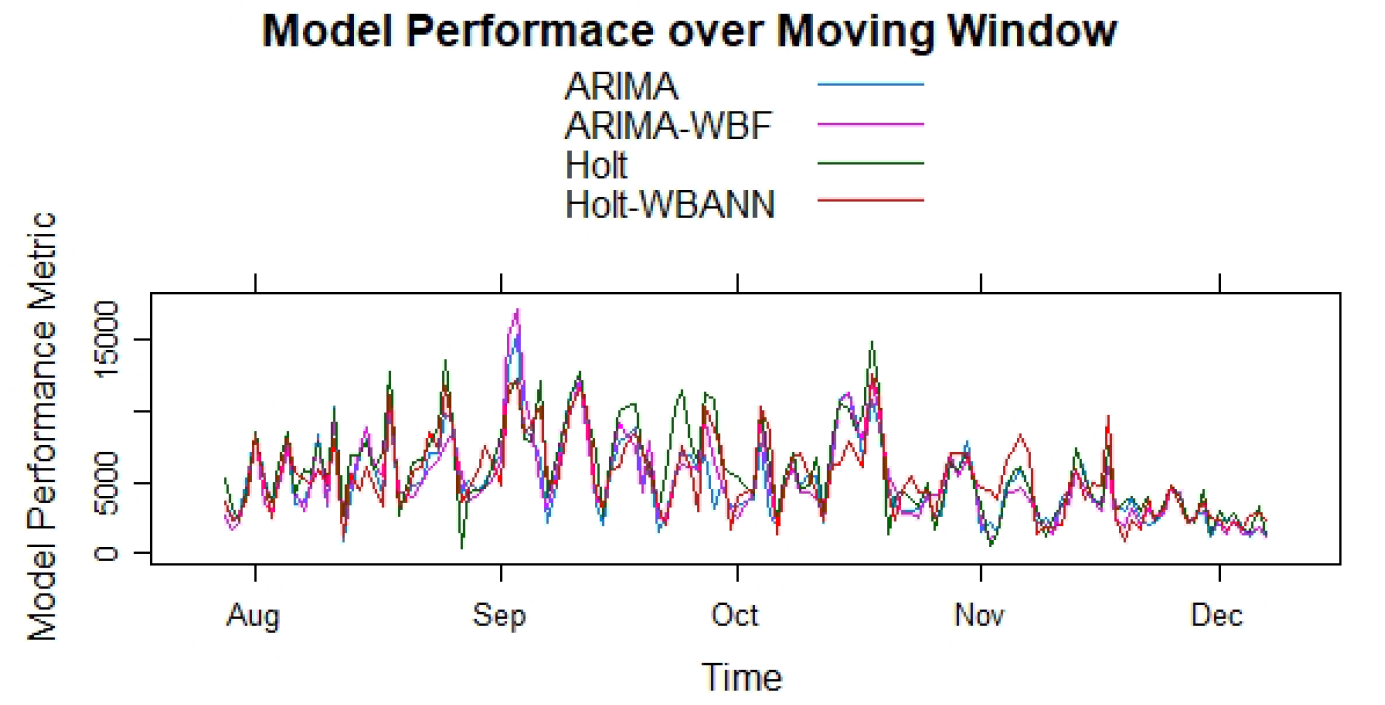"}
\caption{Timeline of moving window performance metric over six hotspot states, showing individual performances of four spotlight models: India}
\label{fig:mw1}
\end{figure}

\begin{figure}[H]
\includegraphics[width=1\linewidth]{"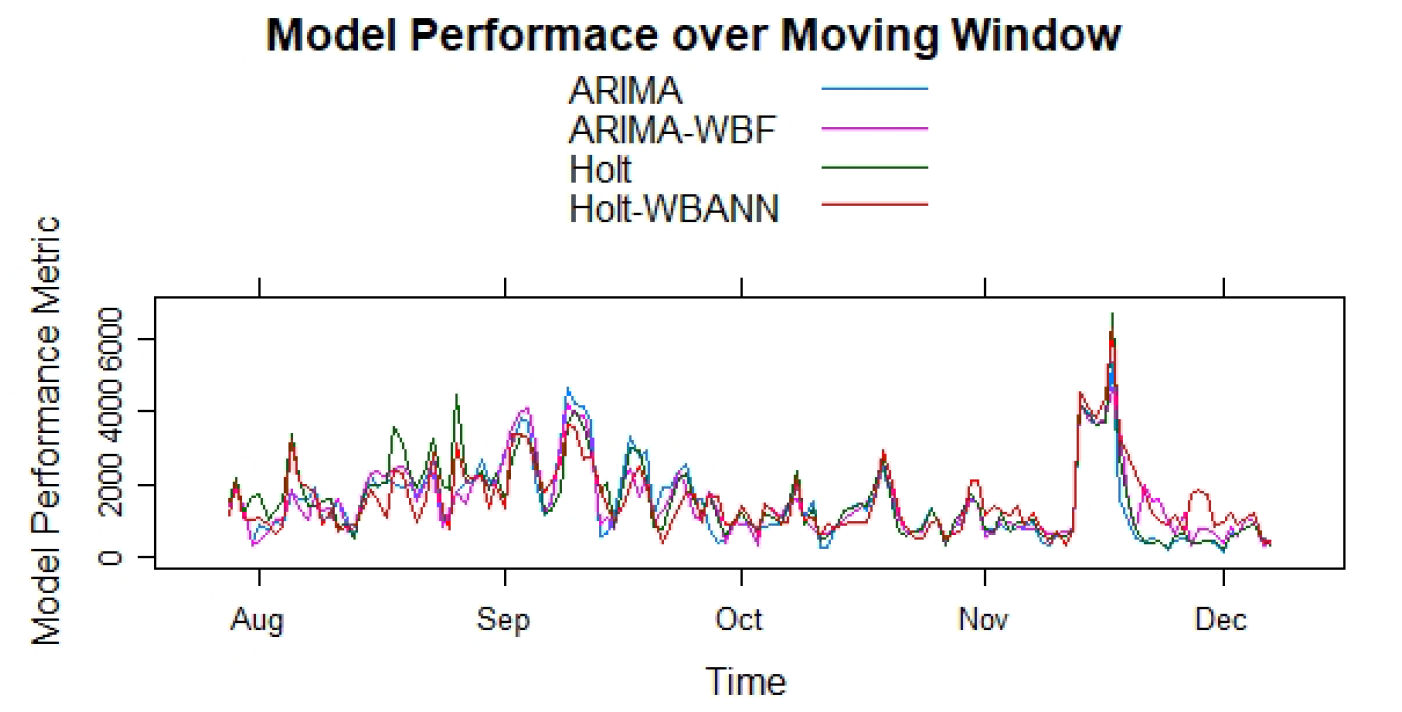"}
\caption{Timeline of moving window performance metric over six hotspot states, showing individual performances of four spotlight models: Maharashtra}
\label{fig:mw2}
\end{figure}

\begin{figure}[H]
\includegraphics[width=1\linewidth]{"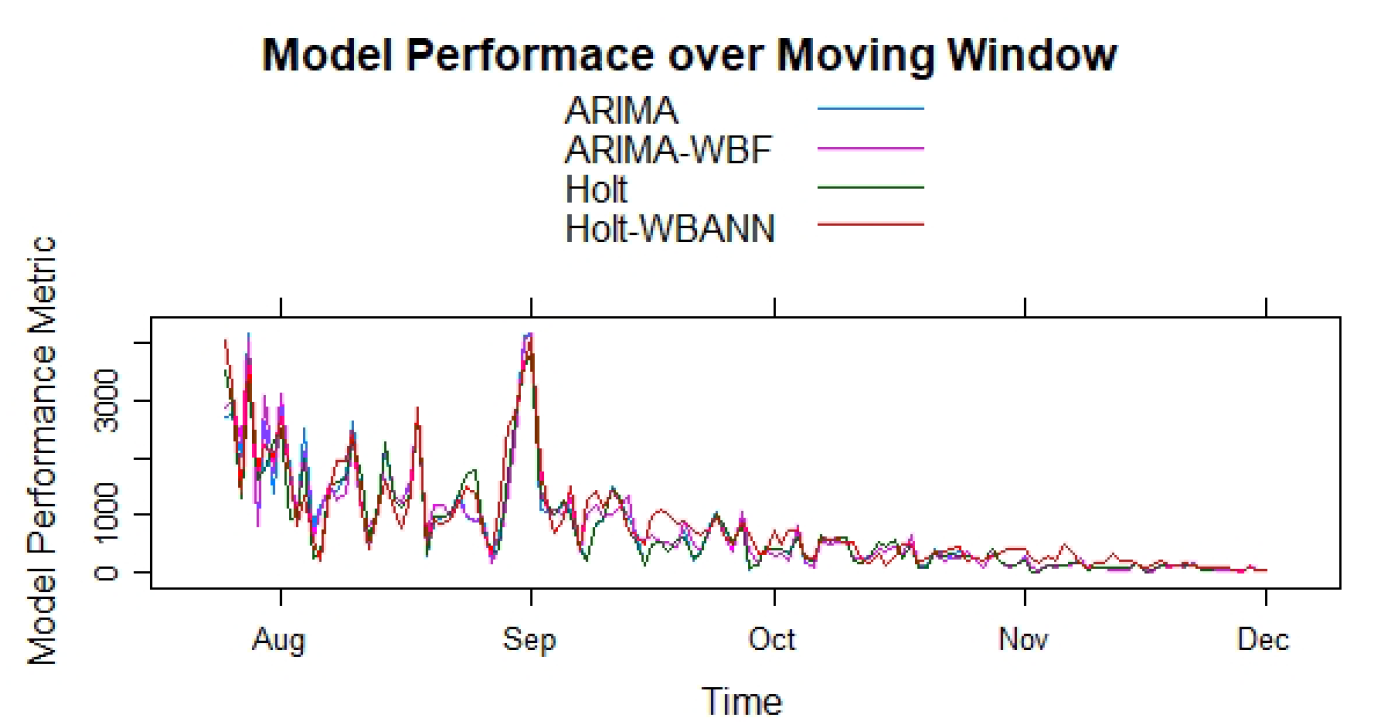"}
\caption{Timeline of moving window performance metric over six hotspot states, showing individual performances of four spotlight models: Andhra Pradesh}
\label{fig:mw3}
\end{figure}

\begin{figure}[H]
\includegraphics[width=1\linewidth]{"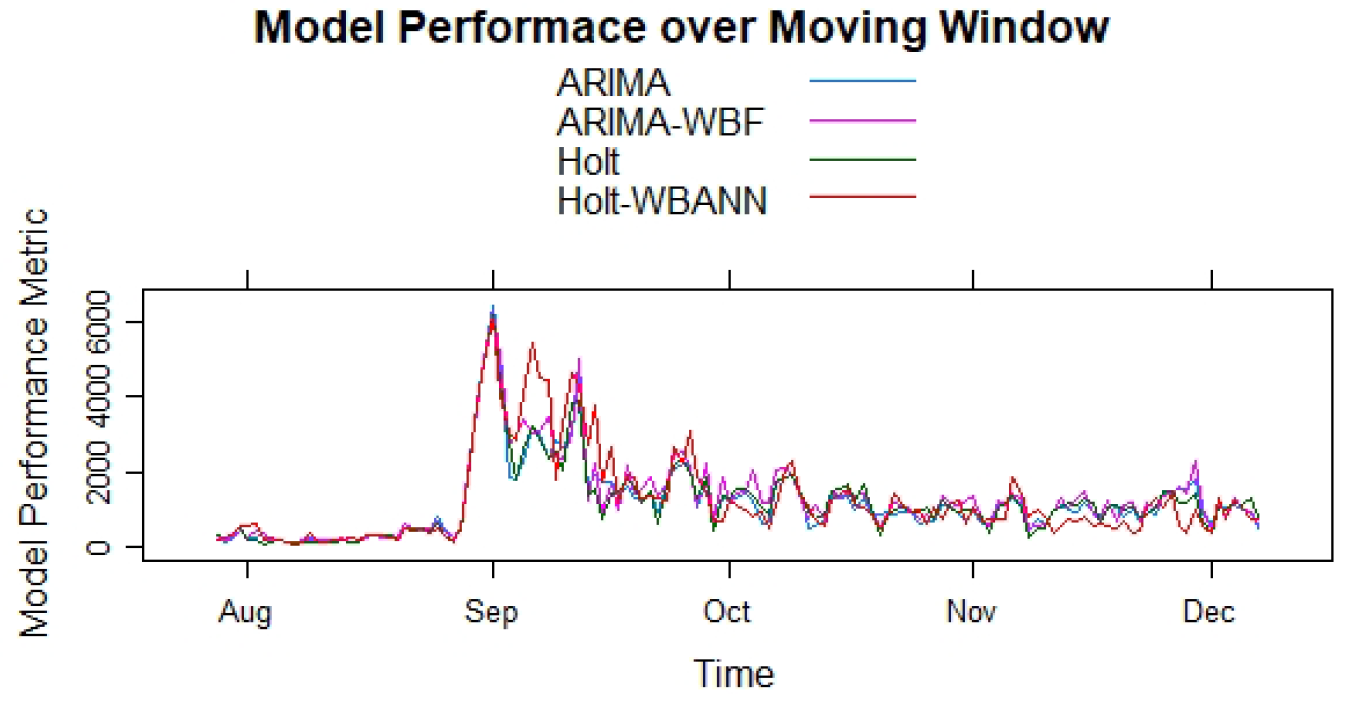"}
\caption{Timeline of moving window performance metric over six hotspot states, showing individual performances of four spotlight models: Karnataka}
\label{fig:mw4}
\end{figure}

\begin{figure}[H]
\includegraphics[width=1\linewidth]{"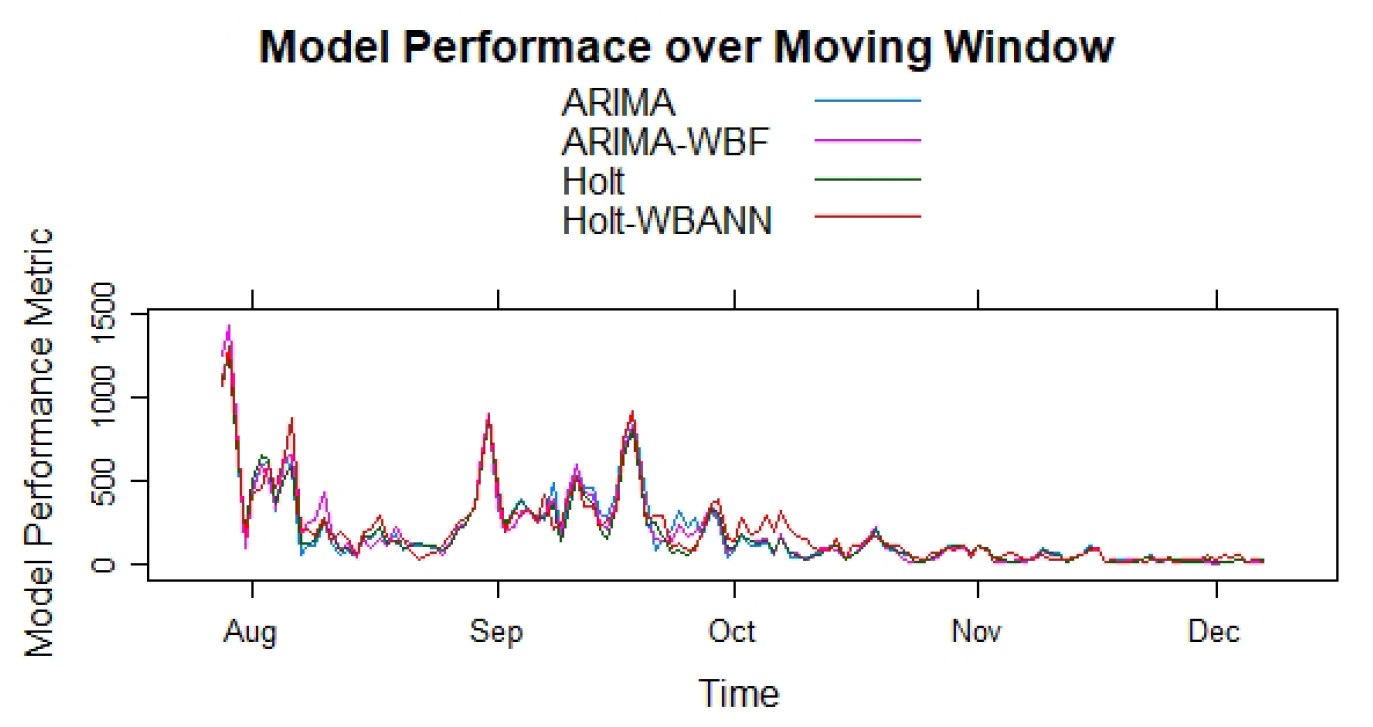"}
\caption{Timeline of moving window performance metric over six hotspot states, showing individual performances of four spotlight models: Tamil Nadu}
\label{fig:mw5}
\end{figure}

\begin{figure}[H]
\includegraphics[width=1\linewidth]{"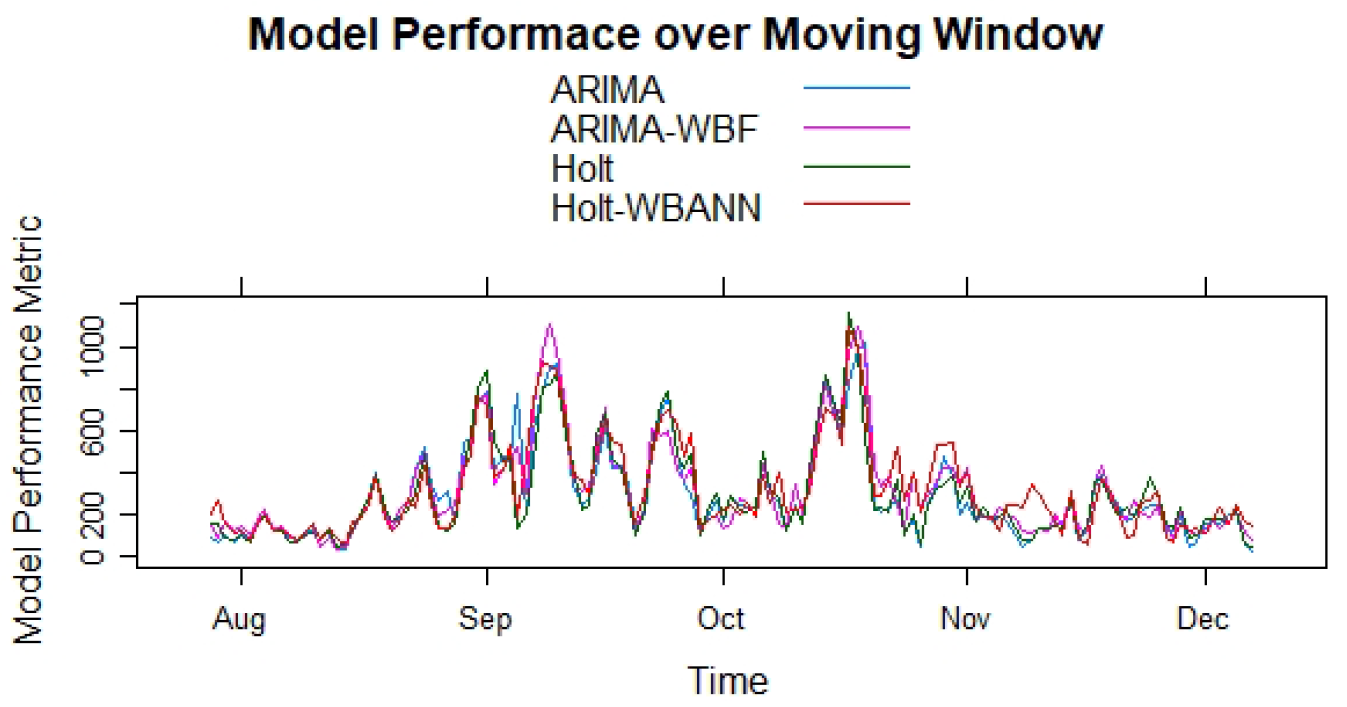"}
\caption{Timeline of moving window performance metric over six hotspot states, showing individual performances of four spotlight models: Chhattisgarh}
\label{fig:mw6}
\end{figure}

\begin{figure}[H]
\includegraphics[width=1\linewidth]{"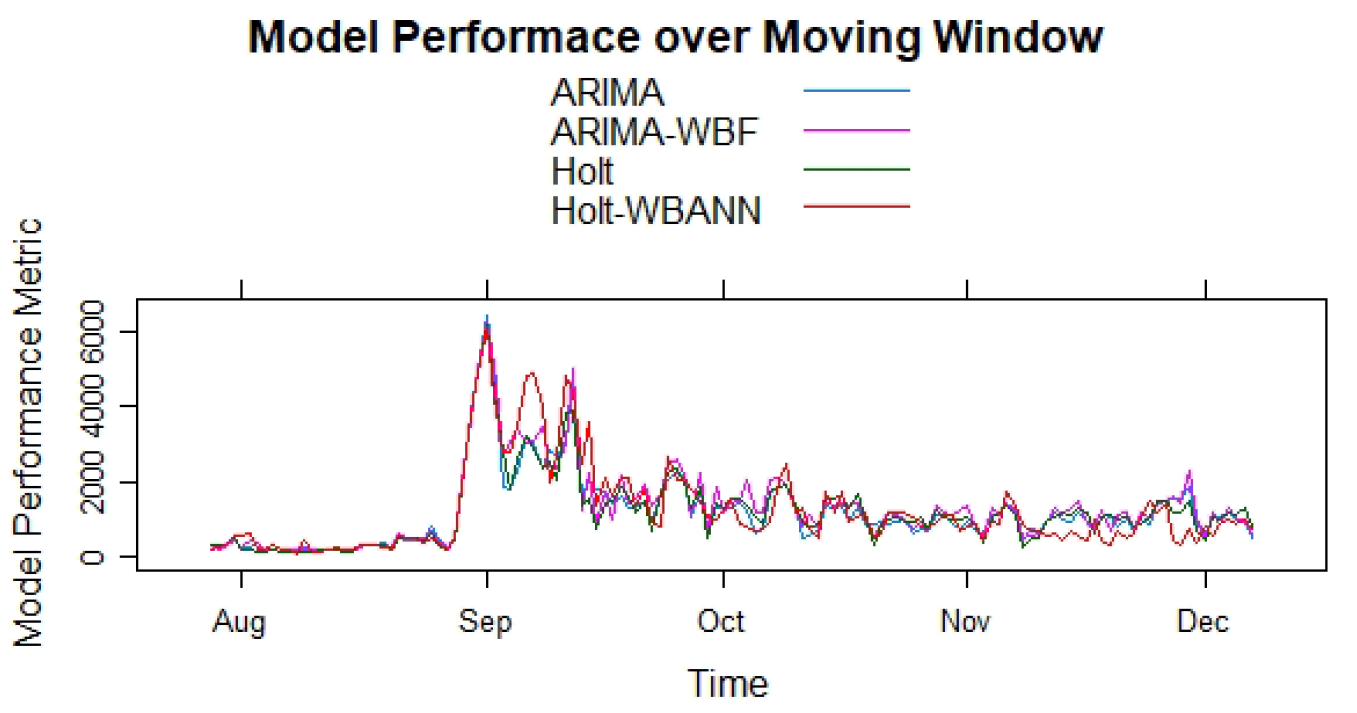"}
\caption{Timeline of moving window performance metric over six hotspot states, showing individual performances of four spotlight models: Kerala}
\label{fig:mw7}
\end{figure}

We have generated forecasts based on an ensemble of primarily four models: The ARIMA, the ARIMA-WBF proposed by \cite{Chakraborty2020}, the Holt's exponential model, and the proposed model. In the latter sections, there have been some discussions on the comparison of our present model with the LSTM model proposed by \cite{Chimmula2020}.The model selection for the final forecast of each state is based on the out of sample prediction metrics over a moving window timeline. The metric used in this paper is $m_j^{(T)}=\frac{{RMSE_j^{(T)}}+{MAE_j^{(T)}}}{2}$, for the $j^{th}$ model $j=1, 2, ..., r$, obtained over multiple windows of $k$ successive time points, $k \in \{1, 2, ..., [\frac{t}{2}]\}$, this metric being plotted over the the discrete time range $T=[\frac{t}{2}]+1, [\frac{t}{2}]+2, ..., (t-k+1)$. Refer to the algorithm provided in \autoref{algo:perf}.\\ Resulting plots haven been provided in \autoref{fig:mw1}, \autoref{fig:mw2}, \autoref{fig:mw3}, \autoref{fig:mw4}, \autoref{fig:mw5}, \autoref{fig:mw6}, \autoref{fig:mw7}.\\
\textbf{Note:} Only out-of-sample performance metrics are taken into consideration.\\

\newpage
\begin{algorithm}[H]
\SetAlgoLined
\begin{enumerate}
\item \textbf{Given:} Multiple Time Series ${y_t}^{(i)}, i=1,2,...,n$, having constant sum such that $Y_t=\sum \limits_{i=1}^n {{y_t}^{(i)}}$. Adjusted forecasts from the previous time point ${\hat{y}_t}^{(i)}$.\\
 \textbf{Input:} Dataset containing daily number of confirmed/deceased cases of individual states, as well as that of the entire country.
\item Generate forecasts ${\hat{y}_{t+1}}^{(1)}, {\hat{y}_{t+1}}^{(2)}$, ..., ${\hat{y}_{t+1}}^{(n)}$  using \textbf{Algorithm 1} for all $n$ states and calculate $\sum \limits_{i=1}^n {\hat{y}_{t+1}}^{(i)}$.
\item Generate forecast $\hat{Y}_{t+1}$ for the univariate time series of the entire country using \textbf{Algorithm 1}. \\
\textbf{If:} $[|Y_t -\hat{Y}_t | \le |\sum \limits_i (y_t^{(i)} - \hat{y}_{t+1}^{(i)})|^{2}]$ : \\
\subitem Assign $\hat{Y}_{t+1} = \sum \limits_i \hat{y}_{t+1}^{(i)}$ \\
\textbf{Else:} \\
\subitem Calculate $d_{t+1}=\hat{Y}_{t+1}-\sum \limits_{i=1}^n {\hat{y}_{t+1}}^{(i)}$.\\
\subitem compute the weights $$w_i =\frac{({{y_t}^{(i)}-{\hat{y}_t}^{(i)})}^2}{{\sum \limits_{i=1}^n {({{y_t}^{(i)}-{\hat{y}_t}^{(i)})}^2}}}, i = 1,2,3,...,n$$.\\
\subitem \textbf{Go to step 4} \\
\item Compute adjusted forecasts ${\hat{y}^{(i)}}_{{t+1}_{corrected}}={\hat{y}_{t+1}}^{(i)}+w_id_{t+1}, i=1,2,...,n$ for each of the $n$ states.
\caption{\textbf{Adjustment factor for sum correction of all states}}
\label{algo:adj}
\end{enumerate}
\end{algorithm}

\begin{algorithm}[H]
\SetAlgoLined
\begin{enumerate}
\item \textbf{Given:} Univariate time series of $t$ observations.\\
\textbf{Input:} Dataset containing daily number of confirmed/deceased cases of any individual state, or that of the entire country.
\item Partition the time series into a training set (here we have considered half of the entire series) and a testing set of size $k$. \\Note that, we have not hampered the natural ordering of the time series. If $[\frac{t}{2}]$ elements are considered as the training set for the first iteration, then the testing set will consist of $([\frac{t}{2}]+1)^{th}, ([\frac{t}{2}]+2)^{th}, ..., ([\frac{t}{2}]+k)^{th}$ elements.
 \caption{\textbf{Model Performance Monitoring}}
\item Train $r$ models based on the training set and generate forecasts $\hat{y}_{[\frac{t}{2}]+1}, \hat{y}_{[\frac{t}{2}]+2}, ..., \hat{y}_{[\frac{t}{2}]+k}$. Compute out of sample residuals and subsequently, the RMSE and the MAE, for all the $r$ models.
\item Compute the metric $m_j^{(T)}=\frac{{RMSE_j^{(T)}}+{MAE_j^{(T)}}}{2}, j=1, 2, ..., r$; $T=[\frac{t}{2}]+1, [\frac{t}{2}]+2, ..., (t-k+1)$.
\item For each point in the timeline $T = [\frac{t}{2}]+1, [\frac{t}{2}]+2, ..., (t-k+1)$, compute $\psi_T=\arg \min \limits_{j} m_j^{(k)};T=[\frac{t}{2}]+1, [\frac{t}{2}]+2, ..., (t-k+1)$.
\item The mode of $\psi_T$ can be considered as the model that gives best out of sample predictions throughout the timeline.
\caption{\textbf{Model Performance Monitoring}}
\label{algo:perf}
\end{enumerate}
\end{algorithm}

\textbf{Remark:} The choice of model performance metric chosen here is quite arbitrary. In many cases, it has been observed that, a model may have a lower RMSE than another model, but has a higher MAE, and vice versa. It may be argued that the choice of RMSE is best when residual values are a little larger (if both models generate residual values less than 1, then the MAE might be a better choice).\\

It may also be worthwhile to note that instead of taking the mode of $\psi_T$, a better alternative is to assign higher weights to recent data points, so as to truly select the best performing model.

\subsection{Estimation of \texorpdfstring{$R_0$}{r\_0}}
The role of the Basic Reproduction Number, $R_0$, in any epidemic is vital for monitoring disease spread and other dynamics as well. It is essentially, the average number of people infected by a single individual in a geographical region. For its estimation, we reiterate the "model agnostic" approach adopted by \cite{das2020prediction}, as it makes sense to use classical epidemiological models for understanding ground reality. As in the work of \cite{das2020prediction} and \cite{Khrapov2020}, we too have used the Susceptible-Infected-Recovered (SIR) model for estimating $R_0$. One may also use compartmental models like the SEIR (where the E stands for "Exposed"), or for that matter, even the SEIPAHRF [Susceptible - Exposed - Infected - (P is for the class of super-spreaders) - Asymptomatic - Hospitalized - Recovered - Fatality] model for even more accurate estimation, as suggested by \cite{ndairou2020mathematical}. \autoref{table:r0} provides the $R_0$ estimates, its 95\% confidence interval(s), as well as the MSE in estimating the mean and shape parameter, $\mu$ and $\kappa$ respectively, as well as the estimates of the parameters themselves, $\hat{\mu}$ and $\hat{\kappa}$, respectively.

\subsection{Model Shelf Life}
\begin{figure}
\includegraphics[width=1\linewidth]{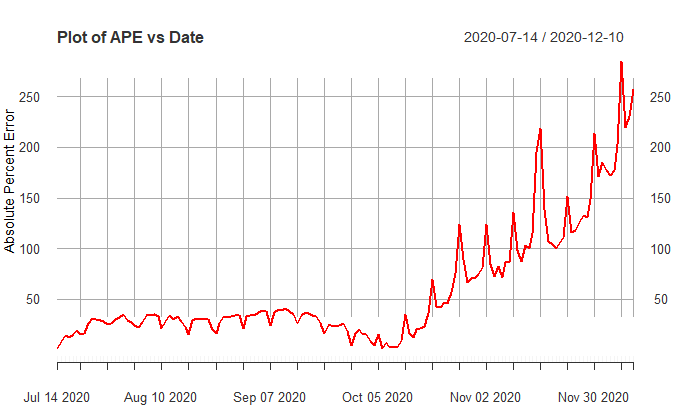}
\caption{Plot of APE(t) over 150 out of sample observations (ordered) for the entire country of India}
\label{fig:ape1}
\end{figure}

\begin{figure}
\includegraphics[width=1\linewidth]{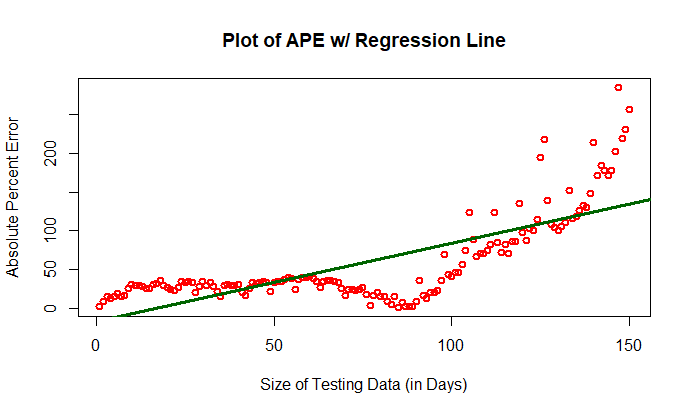}
\caption{Plot of APE(t) over 150 out of sample observations (ordered) for the entire country of India with regression line}
\label{fig:ape2}
\end{figure}

With any human crisis, in this case, a pandemic, it has been seen throughout history that there always comes with it, an abundance of ideas and solutions. The choice of a unique solution on which to focus collectively, becomes tougher with every additional solution present at a point in time. \\ \\
One of the many ways to select the best model for a pandemic situation, typically when the focus is on surveillance or monitoring is to choose the model that holds its predictive power for the longest period of time. In this section, we focus on a very simple methodology to root out the best model among multiple options on the basis of their "longevity". \\ \\
Here, as per custom, we split the time series into two holdout sets, $\{y_1, y_2, ..., y_m\}$ and $\{y_{m+1}, y_{m+2}, ..., y_T \}$. We trained our proposed model on the $m$ training points, and calculated the Absolute Percent Errors for the $(T - m)$ predictions on the testing set.
$$APE(t) = \frac{|y_t - {\hat{y}}_t|}{y_t} \times 100   ;   t = m+1, m+2, ..., T $$
These points were then plotted as may be viewed in \autoref{fig:ape1}. \\ \\
Generally, the APE's increase with time, but in case of the proposed model, it seemed to be fluctuating for low values of $(T - m)$. Finally at $T=272$ and $m=122$, we saw a substantial increasing graph for which we obtained a model shelf life of 22 days, by simply fitting a regression line of $APE(t)$ on $t$, and obtaining the value of $t$ for which $APE(t)$ touched 5\% (see \autoref{fig:ape2}).

\section{Results}
As we can see from \autoref{tab:1}, \autoref{tab:2} and \autoref{tab:4}, the Holt-WBANN model fares exceptionally well when it come to model in-sample fit and  out-of-sample forecast accuracy.\\ 
For the model performance metrics,as mentioned in the \textbf{Section 2.3}, the initial points of reference of time $t$ is $10^{th}$ January, 2021. The width of the moving window considered was 4. The proposed metric here, refers to the percentage of the timeline  dominated by a particular model (By dominated, we mean that the model gives minimum value of $\frac{RMSE+MAE}{2}$ ). \\
For comparison with the ARIMA and the ARIMA-hybrid, model diagnostics included the generation of the ACF and the PACF plots provided in \autoref{tab:3}. \\
\textbf{Note:} The choice of the width of the moving window here, was completely arbitrary. One may use wider moving windows for large enough sample sizes. \\

Comparisons have also been made with modern models, the most popular among which is the Long-Short Term Memory (Vanilla) variant of Recurrent Neural Networks. \cite{Chimmula2020} have published some results related to transmission (in Canada) forecasting using LSTM networks. We have only compared the forecasts of the six aforementioned hotspot states, and that of the entire country. \autoref{tab:lstm} gives us the metrics required for the comparison.\\

\section{Conclusion}
The uncertainty surrounding the Covid-19 pandemic is still unknown. Over the past few months we have seen countries like Spain, Italy and China among others curb the daily number of cases to a large extent. China has even started to reopen schools after declaring Wuhan, the epicentre of the pandemic, free from the airborne virus.\\ \\
The United States still leads the race among the highest count of cases with Russia and India coming a close second and third respectively. Although, the total number of daily cases in India has been inching close to almost one lakh per day, significantly more than that of either the US or Russia. \\ \\
Regardless, forecasting is a vital part of the decision making process, especially considering the global pandemic the Covid-19 has engulfed the world in. In this paper, we have proposed a hybrid Holt/WBANN model which explains the non stationary and non linear present in the dataset. The model is best used for monitoring purpose only. We have also provided one day forecasts for all states and Union territories of India. Since we have presented a real time forecast the model can be updated regularly and predictions can be revised. \\ \\
Also, as per the short term forecasts (7 days) the surge in cases show no sign of slowing down. We have identified 10 hotspot states, namely, Maharashtra,   Andhra Pradesh, Karnataka, Tamil Nadu, Chhattisgarh. We have also added Kerala as a reference.\\ \\
According to the RMSE and MAE performance metrics (Out of sample predictions) we see that our model performs best in the following states but also fares well in others states:
\begin{itemize}
\item Maharashtra
\item Andhra Pradesh
\item Karnataka
\item Tamil Nadu
\item Chhattisgarh
\item Kerala.
\end{itemize}

The hybrid Holt WBANN model performs extremely well for India as well. The addition of the adjustment factor in this model even further accentuates the out of sample forecasting ability of the model as is evident from the output given in the appendix below. Even in comparison with modern computational models like the vanilla LSTM recurrent neural network, it provides better overall accuracy in four out of six hotspot states, although the LSTM fares better while forecasting for the entire country. Here, one may be interested in figuring out how forecasting of larger values using this model may affect the out of sample accuracy; whether this relationship is empirical, is open for debate.\\ \\

The calculation and subsequent plotting of the absolute percent error, showed that the model exceeds the 5\% barrier only after 22 days. Although, repeated model refitting may lead to extended shelf-life as more time passes, it is the relation between shelf-life of a model and the size of the dataset it is trained on which remains to be explored.\\ \\

The estimated $R_0$ based on the generic SIR model developed by \cite{das2020prediction}, seems to portray that the general condition has improved overall in the entire country and most of the other states. However, from the estimates obtained, it only fair to conclude that Andhra Pradesh and Kerala have a higher $R_0$ as compared to the other hotspots.

\begin{center}
\textbf{Acknowledgement} \\
\end{center}
We, the authors are deeply thankful to Mr. Kazi Rahman, Department of Statistics, Sardar Patel University, VVN, Anand, Gujarat, for his contributions in the editing and review of this paper. We are also grateful to the esteemed reviewer(s) without whose inputs and constructive feedback, this article would not have been as interesting.\\

\citeindextrue

\newpage
\section*{Tables and Figures}

\begin{longtable}{|l|l|l|l|}
\caption{Table for comparison of Adjusted and Non-Adjusted values (rounded to the nears unit value) with the true value of number of confirmed cases of Covid-19 as on 10th January, 2021, adjusted values based on data till 9th January, 2021}
\label{tab:adj}\\
\hline
\centering
Location & Unadjusted Value & Adjusted Predicted Value & True Value \\
\hline
India & 16055 & 17685 & 16085 \\
\hline
Andaman and Nicobar Islands & 2 & 1 & 0 \\
\hline
Andhra Pradesh & 210 & 252 & 227 \\
\hline
Arunachal Pradesh & 4 & 9 & 5 \\
\hline
Assam & 17 & 23 & 25 \\
\hline
Bihar & 354 & 333 & 359 \\
\hline
Chandigarh & 53 & 56 & 44 \\
\hline
Chhattisgarh & 863 & 773 & 661 \\
\hline
Delhi	 & 479 & 464 & 399 \\
\hline
Dadra and Nagar Haveli & 1 & 1 & 0 \\
\hline
Goa & 77 & 87 & 66 \\
\hline
Gujarat & 674 & 679 & 671 \\
\hline
Himachal Pradesh & 52 & 34 & 86 \\
\hline
Haryana & 289 & 282 & 234 \\
\hline
Jharkhand & 213 & 221 & 145 \\
\hline
Jammu and Kashmir & 127 & 127 & 113 \\
\hline
Karnataka & 709 & 746 & 792 \\
\hline
Kerala & 5046 & 5084 & 4545 \\
\hline
Ladakh & 16 & 8 & 10 \\
\hline
Lakshadweep & 0 & 0 & 0 \\
\hline
Maharashtra & 4124 & 3903 & 3558 \\
\hline
Meghalaya & 58 & 44 & 19 \\
\hline
Manipur & 41 & 49 & 14 \\
\hline
Madhya Pradesh & 536 & 537 & 620 \\
\hline
Mizoram & 11 & 9 & 17 \\
\hline
Nagaland & 11 & 10 & 3 \\
\hline
Odisha & 187 & 207 & 260 \\
\hline
Punjab & 263 & 263 & 299 \\
\hline
Puducherry & 31 & 34 & 30 \\
\hline
Rajasthan & 439 & 412 & 475 \\
\hline
Sikkim & 0 & 3 & 4 \\
\hline
Telangana & 289 & 335 & 351 \\
\hline
Tamil Nadu & 763 & 752 & 724 \\
\hline
Tripura & 3 & 3 & 8 \\
\hline
Unidentified & 5 & 5 & 0 \\
\hline
Uttar Pradesh & 1074 & 978 & 275 \\
\hline
Uttarakhand & 165 & 133 & 223 \\
\hline
West Bengal & 754 & 867 & 823 \\
\hline
\end{longtable}
\newpage

\begin{table} \small%
\caption{Model performance metrics of different forecasting models for various hotspot states and the entire country of India}
\label{tab:1}
\centering
\begin{adjustbox}{width=1.35\textwidth}
\small
\begin{tabular}{|p{3.1cm}|p{2.4cm}|p{2cm}|p{1.7cm}|p{2.3cm}| p{1.7cm}| p{2.3cm}|}
\hline
Location & Accuracy type & Performance Metrics & ARIMA & ARIMA/WBF & Holt & Holt/WBANN \\
\hline
INDIA & In sample	 & RMSE & 3514.465 & 3587.664 & 3738.819 & 285.983 \\
& &	MAE	 & 2181.369 & 2209.726 & 2263.562 & 202.954 \\
\cline{3-3} \cline{4-4} \cline{5-5} \cline{6-6} \cline{7-7}
& Out of sample & RMSE & 9845.041 & 8593.190 & 9857.528 & 8462.758 \\
& & MAE & 9222.588 & 7882.294 & 9200.756 & 7463.432 \\
\hline
MAHARASHTRA  & In sample & RMSE & 1066.4276 & 1130.8579 & 1100.8528 & 153.1286 \\
& & MAE & 676.5241 & 754.2184 & 696.2935 & 102.7917 \\
\cline{3-3} \cline{4-4} \cline{5-5} \cline{6-6} \cline{7-7}
& Out of sample & RMSE & 2045.250 & 2011.045 & 1983.532 & 2120.690 \\
& & MAE & 951.4600 & 947.4375 & 884.2347 & 907.8484\\
\hline
ANDHRA PRADESH & In sample & RMSE & 625.06805 & 677.98484 & 625.10412 & 37.26153 \\
& & MAE & 308.72666 & 345.79287 & 308.75294 & 22.45116 \\
\cline{3-3} \cline{4-4} \cline{5-5} \cline{6-6} \cline{7-7}
& Out of sample & RMSE & 116.1670 & 149.1013 & 116.9986 &409.2144 \\
& & MAE & 98.36191 & 124.11440 & 99.87684 & 288.87685 \\
\hline
KARNATAKA & In sample & RMSE & 589.51022 & 623.71399 & 620.74529 & 37.78497 \\
& & MAE & 329.48078 & 399.30299 & 338.01977 & 20.01929 \\
\cline{3-3} \cline{4-4} \cline{5-5} \cline{6-6} \cline{7-7}
&Out of sample & RMSE & 295.0593 & 290.6018 & 353.9216 &574.7829 \\
& & MAE & 242.2907 & 232.5452 & 308.2002 & 447.9411 \\
\hline
TAMIL NADU & In sample & RMSE & 174.47032 & 187.40614 & 176.31538 & 28.16981 \\
& & MAE & 106.62399 & 119.53075 & 107.35543 & 19.02123 \\
\cline{3-3} \cline{4-4} \cline{5-5} \cline{6-6} \cline{7-7}
& Out of sample & RMSE & 292.4827 & 276.2654 & 246.7395 & 210.4720 \\
& & MAE & 257.0411 & 240.4029	& 217.5928 & 181.7270 \\
\hline
CHHATTISGARH & In sample & RMSE & 196.4709 & 213.8806 & 205.6074 & 25.0906 \\
& & MAE & 105.89080 & 121.47914 & 107.21559 & 16.78807 \\
\cline{3-3} \cline{4-4} \cline{5-5} \cline{6-6} \cline{7-7}
& Out of sample & RMSE & 472.0904 & 468.8753 & 498.577 & 451.5344 \\
& & MAE & 418.0870 & 416.0293 & 438.0818 & 397.0037 \\
\hline
KERALA & In sample & RMSE & 912.29202 & 1025.59250 & 989.57691 & 49.12563 \\
& & MAE & 415.25064 & 467.01815 & 454.18426 & 28.81994 \\
\cline{3-3} \cline{4-4} \cline{5-5} \cline{6-6} \cline{7-7}
& Out of sample & RMSE & 958.2916 & 1006.5044 & 1106.8631 & 909.4099 \\
&  & MAE & 777.3032 & 802.5384 & 811.2630 & 700.1036 \\
\hline
\end{tabular}
\end{adjustbox}
\end{table}

\begin{table} \small%
\caption{Model performance across hotspot states and the entire country, measured using proposed performance metric for four spotlight models}
\label{tab:2}
\centering
\begin{adjustbox}{width=1.35\textwidth}
\small
\begin{tabular}{|p{3.1cm}|p{3cm}|p{5cm}|}
\hline
LOCATION & MODEL & PROPOSED PERFORMANCE METRIC \\
\hline
INDIA & ARIMA & 23.30 \\
\cline{2-2} \cline{3-3}
& ARIMA-WBF & 34.58 \\
\cline{2-2} \cline{3-3}
& HOLT & 10.52 \\
\cline{2-2} \cline{3-3}
& HOLT-WBANN & 31.57 \\
\hline
MAHARASHTRA & ARIMA & 26.31 \\
\cline{2-2} \cline{3-3}
& ARIMA-WBF & 23.30 \\
\cline{2-2} \cline{3-3}
& HOLT & 16.54 \\
\cline{2-2} \cline{3-3}
& HOLT-WBANN & 33.83 \\
\hline
ANDHRA PRADESH & ARIMA & 16.15 \\
\cline{2-2} \cline{3-3}
& ARIMA-WBF & 39.23 \\
\cline{2-2} \cline{3-3}
& HOLT & 22.30\\
\cline{2-2} \cline{3-3}
& HOLT-WBANN & 22.30 \\
\hline
KARNATAKA & ARIMA & 24.06 \\
\cline{2-2} \cline{3-3}
& ARIMA-WBF &18.79 \\
\cline{2-2} \cline{3-3}
& HOLT & 29.32 \\
\cline{2-2} \cline{3-3}
& HOLT-WBANN & 27.81 \\
\hline
TAMIL NADU & ARIMA & 26.31 \\
\cline{2-2} \cline{3-3}
& ARIMA-WBF & 25.56 \\
\cline{2-2} \cline{3-3}
& HOLT & 21.80 \\
\cline{2-2} \cline{3-3}
& HOLT-WBANN & 26.31 \\
\hline
CHHATTISGARH & ARIMA & 24.81 \\
\cline{2-2} \cline{3-3}
& ARIMA-WBF & 19.54 \\
\cline{2-2} \cline{3-3}
& HOLT & 29.32 \\
\cline{2-2} \cline{3-3}
& HOLT-WBANN & 26.31 \\
\hline
KERALA & ARIMA & 26.31 \\
\cline{2-2} \cline{3-3}
& ARIMA-WBF & 9.77 \\
\cline{2-2} \cline{3-3}
& HOLT & 23.30 \\
\cline{2-2} \cline{3-3}
& HOLT-WBANN & 40.60 \\
\hline
\end{tabular}
\end{adjustbox}
\end{table}

\begin{table} \small%
\caption{Model performance metrics of proposed forecasting model and LSTM model for various hotspot states and the entire country of India}
\label{tab:lstm}
\centering
\begin{adjustbox}{width=1.35\textwidth}
\small
\begin{tabular}{|p{3.1cm}|p{2.4cm}|p{2cm}|p{1.7cm}|p{1.7cm}|}
\hline
Location & Accuracy type & Performance Metrics & LSTM & Holt/ WBANN \\
\hline
INDIA & In sample	 & RMSE & 108.911  & 285.983 \\
& &	MAE	 & 194.141 & 202.954 \\
\cline{3-3} \cline{4-4} \cline{5-5}
& Out of sample & RMSE & 8118.554 & 8462.758 \\
& & MAE & 6995.313 & 7463.432 \\
\hline
MAHARASHTRA  & In sample & RMSE & 304.082 & 153.1286 \\
& & MAE & 284.9138 & 102.7917 \\
\cline{3-3} \cline{4-4} \cline{5-5}
& Out of sample & RMSE & 1794.813 & 2120.690 \\
& & MAE & 805.2351 & 907.8484\\
\hline
ANDHRA PRADESH & In sample & RMSE & 301.04143 & 37.26153 \\
& & MAE & 228.359 & 22.45116 \\
\cline{3-3} \cline{4-4} \cline{5-5}
& Out of sample & RMSE & 104.4142 &409.2144 \\
& & MAE & 93.831 & 288.87685 \\
\hline
KARNATAKA & In sample & RMSE & 41.91323 & 37.78497 \\
& & MAE & 29.19837 & 20.01929 \\
\cline{3-3} \cline{4-4} \cline{5-5}
&Out of sample & RMSE & 610.5571 &574.7829 \\
& & MAE & 530.6154 & 447.9411 \\
\hline
TAMIL NADU & In sample & RMSE & 21.130502 & 28.16981 \\
& & MAE & 14.93965 & 19.02123 \\
\cline{3-3} \cline{4-4} \cline{5-5}
& Out of sample & RMSE & 280.0781 & 210.4720 \\
& & MAE & 241.1173 & 181.7270 \\
\hline
CHHATTISGARH & In sample & RMSE & 29.8732 & 25.0906 \\
& & MAE & 18.2175 & 16.78807 \\
\cline{3-3} \cline{4-4} \cline{5-5}
& Out of sample & RMSE & 481.1166 & 451.5344 \\
& & MAE & 423.6472 & 397.0037 \\
\hline
KERALA & In sample & RMSE & 733.4561 & 49.12563 \\
& & MAE & 323.19537 & 28.81994 \\
\cline{3-3} \cline{4-4} \cline{5-5}
& Out of sample & RMSE & 1105.6100 & 909.4099 \\
&  & MAE & 791.7374 & 700.1036 \\
\hline
\end{tabular}
\end{adjustbox}
\end{table}

\begin{table}
\caption{Training datasets and corresponding ACF, PACF plots of Confirmed Cases for various hotspot states and India}
\label{tab:3}
\centering
\begin{adjustbox}{width=1.3\textwidth}
\small
\begin{tabular} {|l|l|l|l|}
\hline
Location & Training Data & ACF & PACF\\
\hline\\
Maharashtra & \includegraphics[width=0.3\linewidth]{"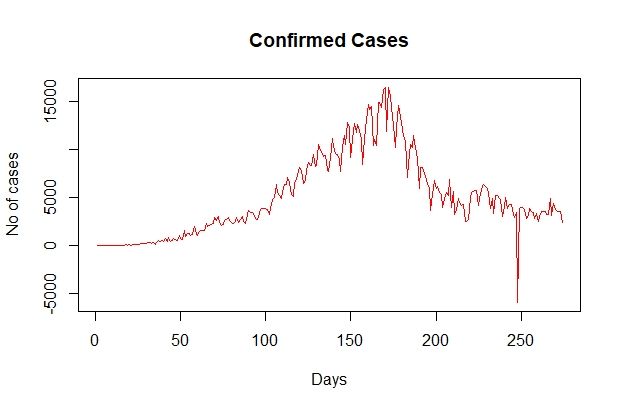"}
\label{fig:mh-tr} & \includegraphics[width=0.3\linewidth]{"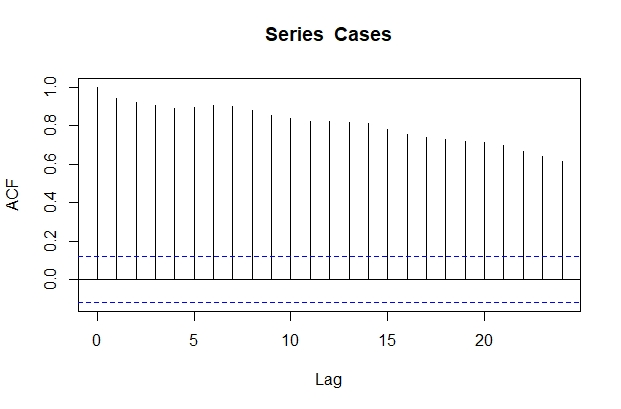"}
\label{fig:mh-acf} & \includegraphics[width=0.3\linewidth]{"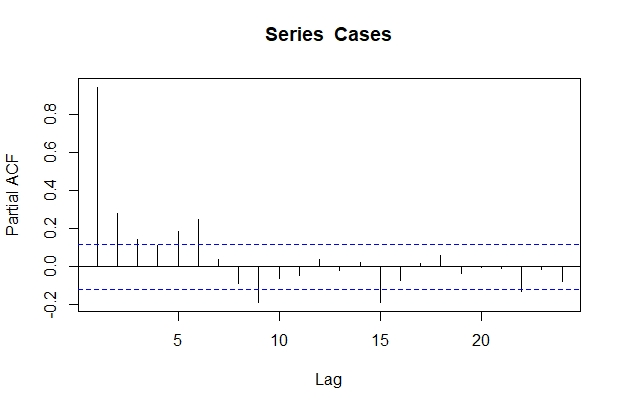"}
\label{fig:mh-pacf}\\
\hline\\
Andhra Pradesh & \includegraphics[width=0.3\linewidth]{"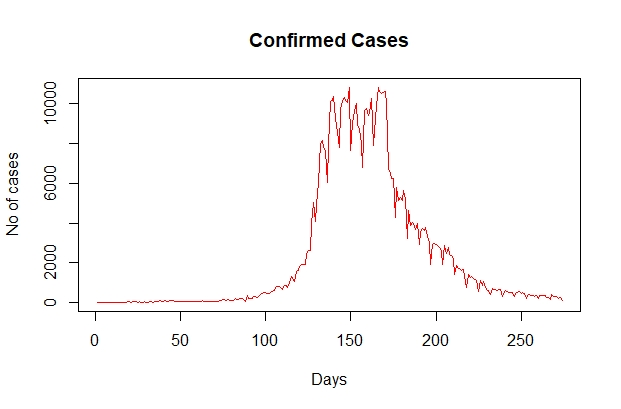"}
\label{fig:ap-tr} & \includegraphics[width=0.3\linewidth]{"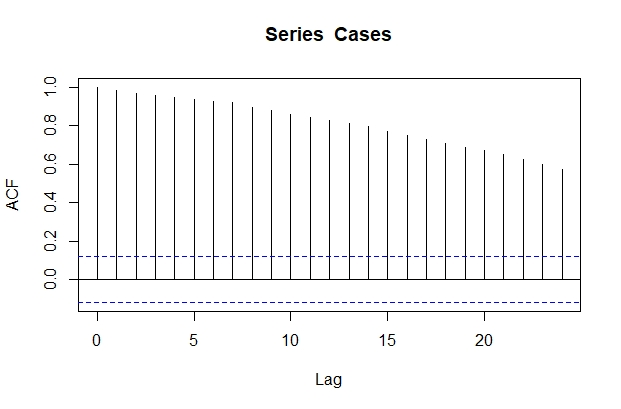"}
\label{fig:ap-acf} & \includegraphics[width=0.3\linewidth]{"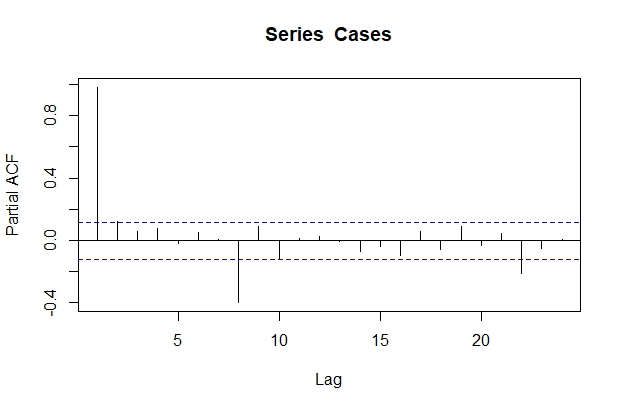"}
\label{fig:ap-pacf}\\
\hline\\
Karnataka & \includegraphics[width=0.3\linewidth]{"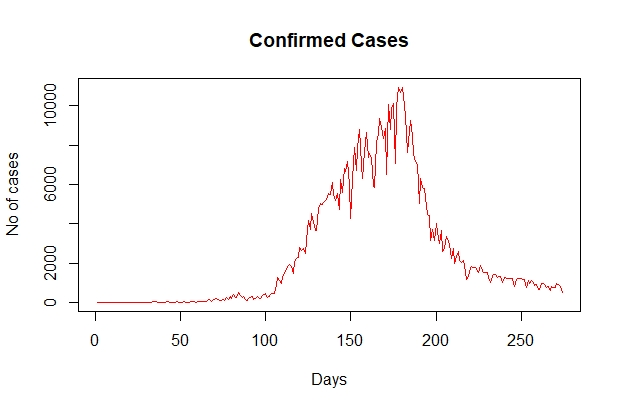"}
\label{fig:ka-tr} & \includegraphics[width=0.3\linewidth]{"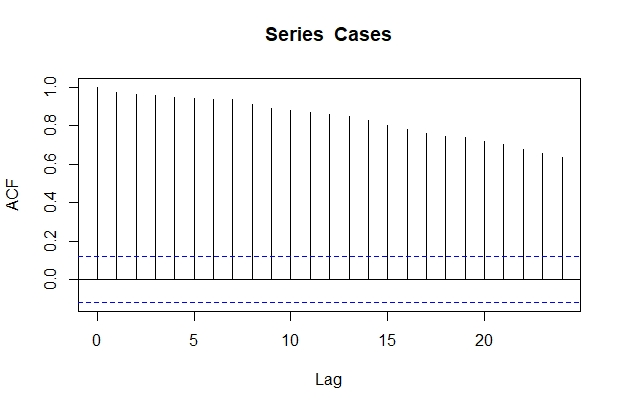"}
\label{fig:ka-acf} & \includegraphics[width=0.3\linewidth]{"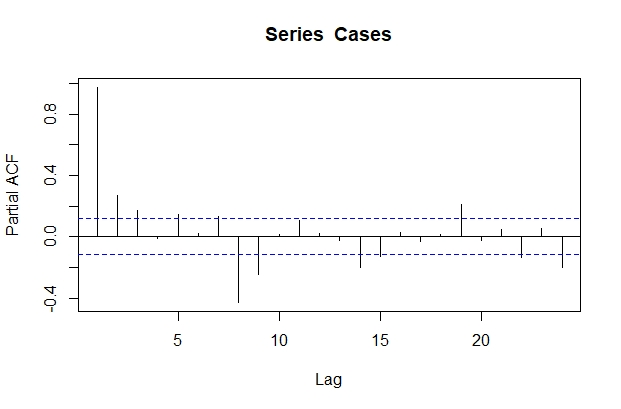"}
\label{fig:ka-pacf}\\
\hline\\
Tamil Nadu & \includegraphics[width=0.3\linewidth]{"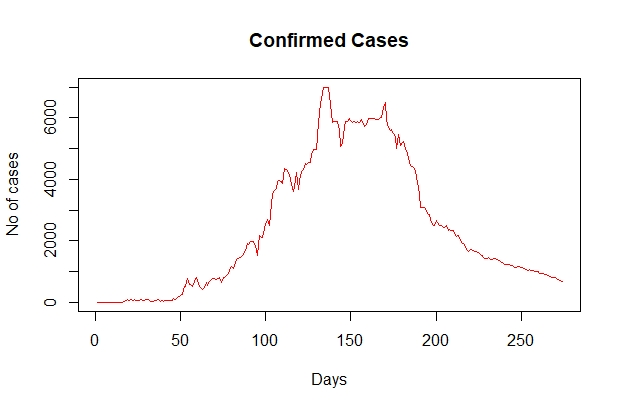"}
\label{fig:tn-tr} & \includegraphics[width=0.3\linewidth]{"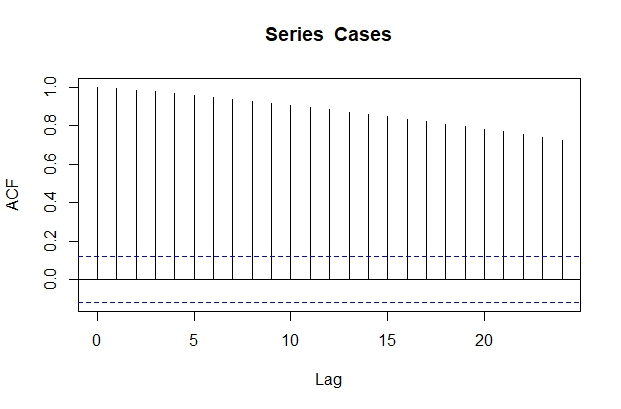"}
\label{fig:tn-acf} & \includegraphics[width=0.3\linewidth]{"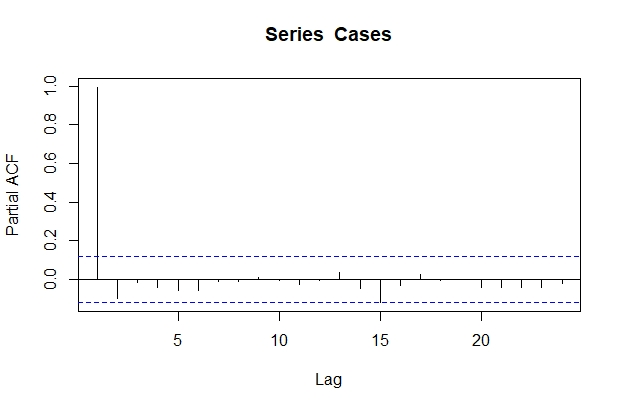"}
\label{fig:tn-pacf}\\
\hline\\
Chhattisgarh & \includegraphics[width=0.3\linewidth]{"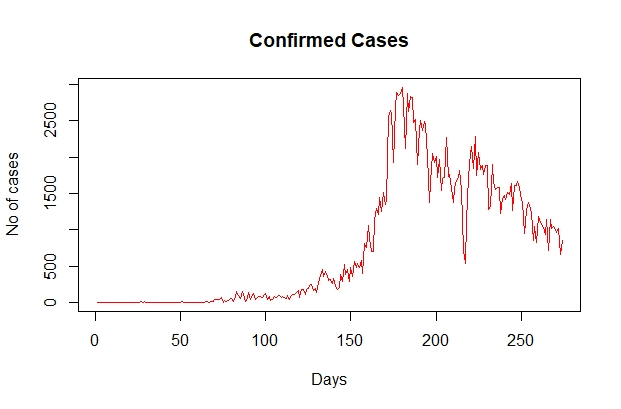"}
\label{fig:ct-tr} & \includegraphics[width=0.3\linewidth]{"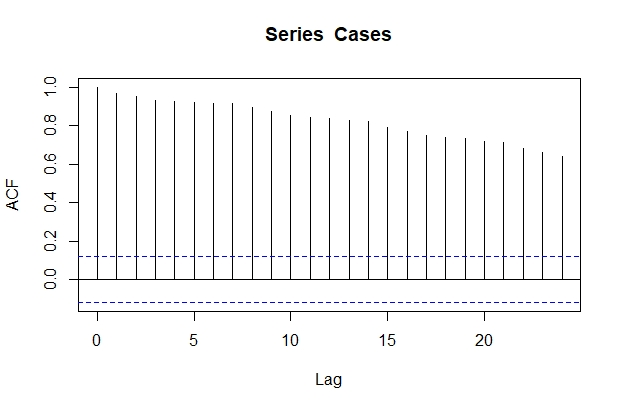"}
\label{fig:ct-acf} & \includegraphics[width=0.3\linewidth]{"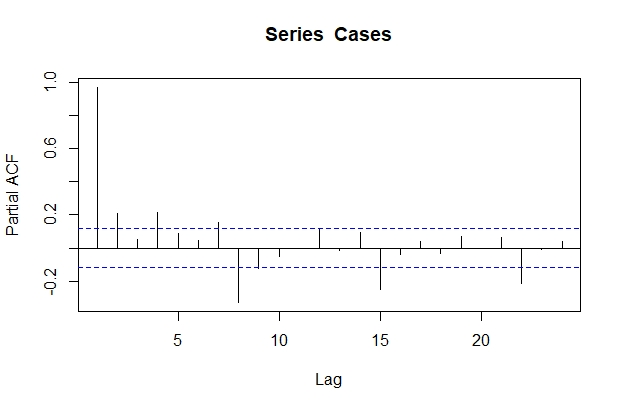"}
\label{fig:ct-pacf}\\
\hline\\
Kerela & \includegraphics[width=0.3\linewidth]{"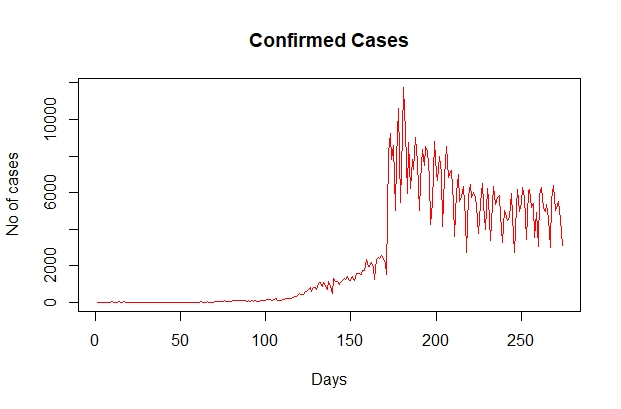"}
\label{fig:kl-tr} & \includegraphics[width=0.3\linewidth]{"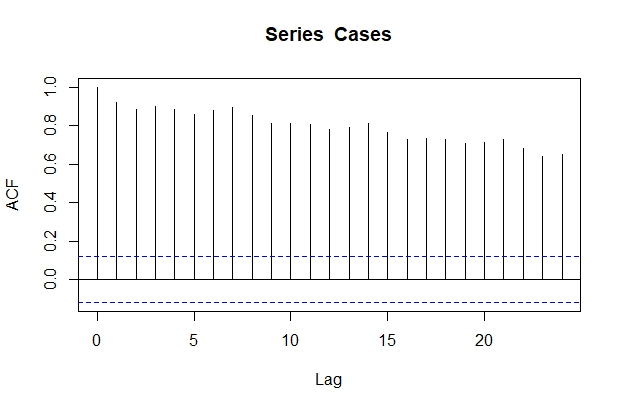"}
\label{fig:kl-acf} & \includegraphics[width=0.3\linewidth]{"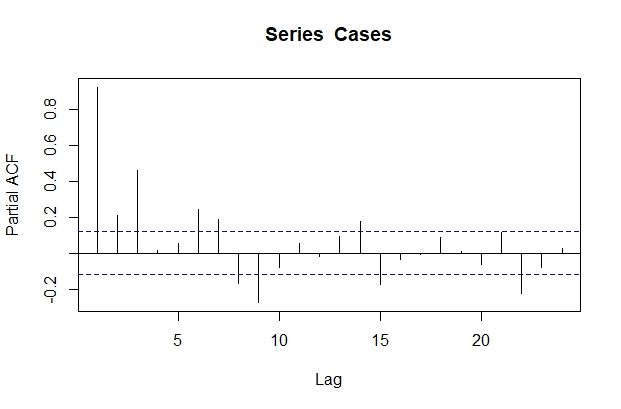"}
\label{fig:kl-pacf}\\
\hline
\end{tabular}
\end{adjustbox}
\end{table}

\begin{table}
\caption{In sample and Out of sample fits for various hotspot states and India}
\label{tab:4}
\centering
\begin{adjustbox}{width=1.1\textwidth}
\small
\begin{tabular} {|l|l|l|}
\hline \\
Location & In sample fits & Out of sample fits \\
\hline \\
India & \includegraphics[width=0.3\linewidth]{"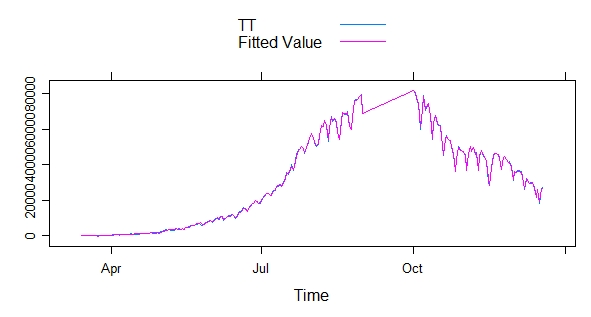"}
\label{fig:tt-in} & \includegraphics[width=0.3\linewidth]{"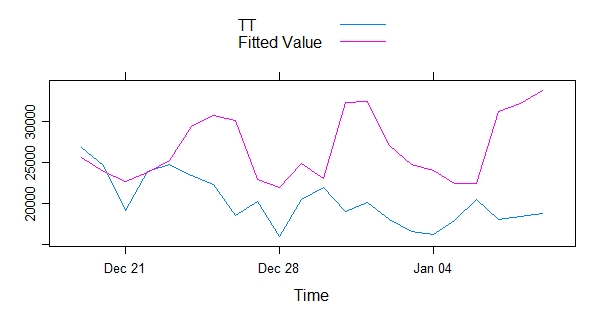"}
\label{fig:tt-out} \\
\hline \\
Maharashtra & \includegraphics[width=0.3\linewidth]{"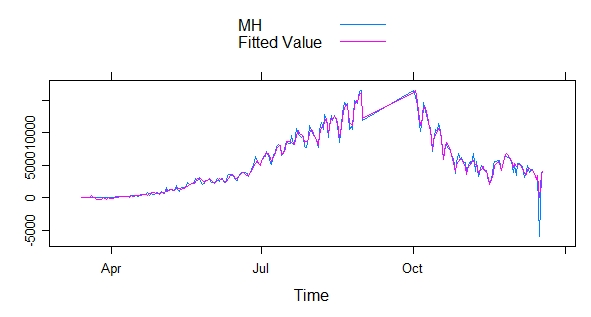"}
\label{fig:mh-in} & \includegraphics[width=0.3\linewidth]{"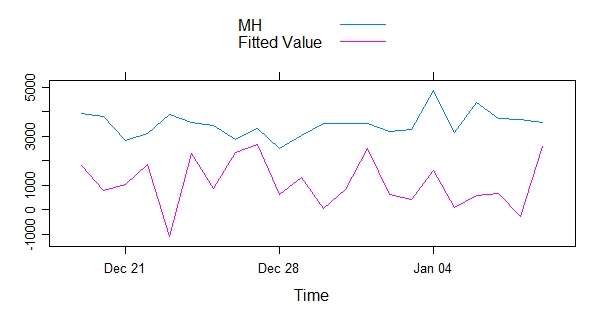"}
\label{fig:mh-out}\\
\hline \\
Andhra Pradesh & \includegraphics[width=0.3\linewidth]{"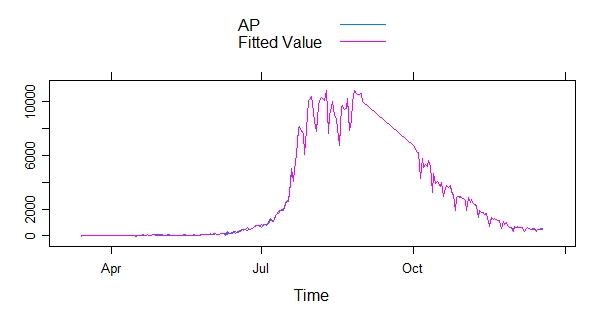"}
\label{fig:ap-in} & \includegraphics[width=0.3\linewidth]{"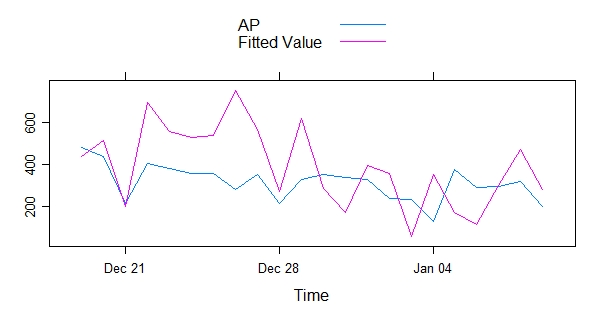"}
\label{fig:ap-out} \\
\hline \\
Karnataka & \includegraphics[width=0.3\linewidth]{"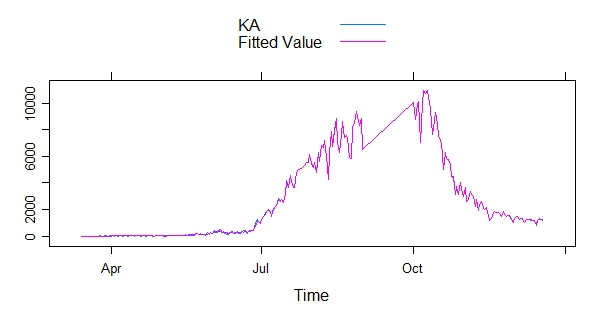"}
\label{fig:ka-in} & \includegraphics[width=0.3\linewidth]{"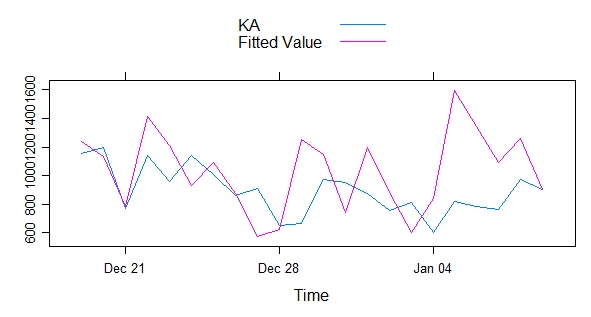"}
\label{fig:ka-out} \\
\hline \\
Tamil Nadu & \includegraphics[width=0.3\linewidth]{"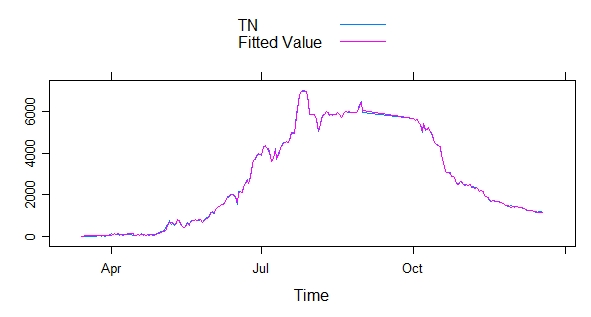"}
\label{fig:tn-in} & \includegraphics[width=0.3\linewidth]{"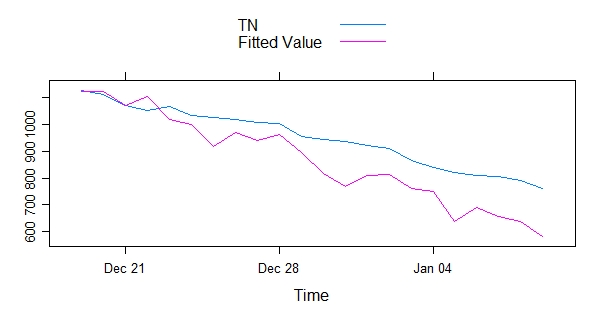"}
\label{fig:tn-out} \\
\hline \\
Chhattisgarh & \includegraphics[width=0.3\linewidth]{"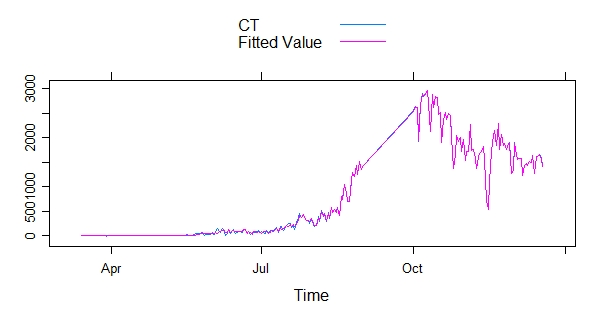"}
\label{fig:ct-in} & \includegraphics[width=0.3\linewidth]{"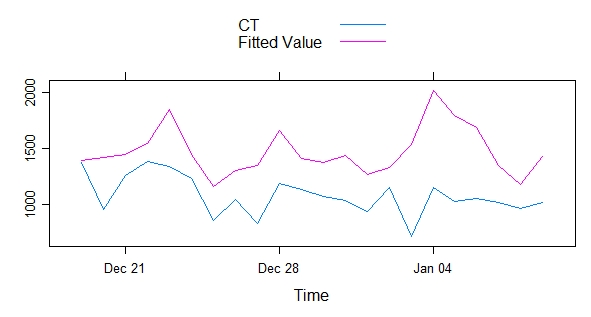"}
\label{fig:ct-out}\\
\hline \\
Kerela & \includegraphics[width=0.3\linewidth]{"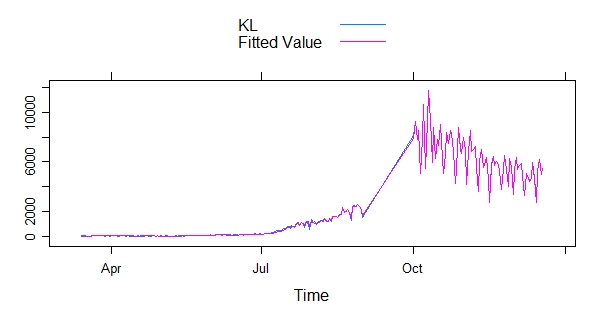"}
\label{fig:kl-in} & \includegraphics[width=0.3\linewidth]{"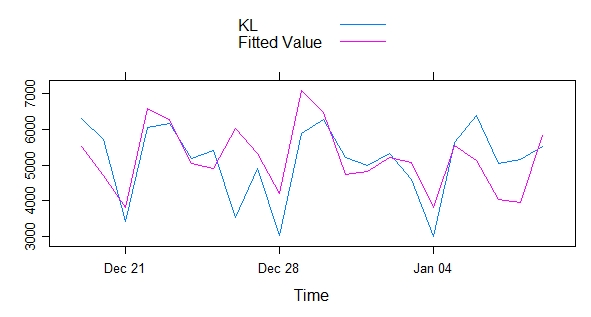"}
\label{fig:kl-out}\\
\hline
\end{tabular}
\end{adjustbox}
\end{table}

\begin{table}
\small%
\caption{$R_0$ estimates for various hotspot states and the entire country of India}
\label{table:r0}
\centering
\begin{adjustbox}{width=1.35\textwidth}
\small
\begin{tabular}{|p{2.6cm}|p{1cm}|p{1cm}|p{2cm}|p{2cm}| p{2cm}| p{2cm}|}
\hline
	&Mean 	&Shape	&$R_0$	&$R_0^{lower}$	&$R_0^{upper}$ &MSE \\
\hline
India	&0.1	&10	&1.400733	&1.400396	&1.40107	&2.15545$\times 10^9$ \\
\hline
Maharashtra	&0.1	&10	&1.374005892	&1.373260312	&1.374751675	&82521981.68 \\
\hline
Andhra Pradesh	&0.1	&10	&1.646931643	&1.645252755	&1.648612205	&21897999.65 \\
\hline
Karnataka	&0.1	&10	&1.381081809	&1.38020343	&1.381960626	&21740034.76 \\
\hline
Tamil Nadu	&0.7	&10	&1.473472026	&1.471648288	&1.475297445	&11918982.56 \\
\hline
Chhattisgarh	&9.1	&8.8	&1.439878417	&1.437038795	&1.442730703	&1180444.848 \\
\hline
Kerala	&0.1	&10	&1.543459206	&1.541776509	&1.545143891	&4678754.632 \\
\hline
\end{tabular}
\end{adjustbox}
\end{table}

\newpage
\noindent
{\bf Agniva Das} \\
Department of Statistics, The Maharaja Sayajirao University of Baroda\\
Vadodara - 390002\\
E-mail: agniva.d-statphd@msubaroda.ac.in\\
GitHub link for the R codes for the models and plots in this paper: \href{https://github.com/dasagniva/wbann}{https://github.com/dasagniva/wbann} \\
YouTube channel link:\href{https://www.youtube.com/channel/UCb5GdwgdvPClALVv9I0avKw}{https://www.youtube.com/channel/UCb5GdwgdvPClALVv9I0avKw} \\
\vspace{.1in}

\noindent
{\bf Dr. Kunnummal Muralidharan}\\
Dept. of Statistics, The Maharaja Sayajirao University of Baroda\\
Vadodara - 390002\\
E-mail: lmv\_murali@yahoo.com\\

\end{document}